\documentclass{article}
\usepackage{spconf,amsmath,graphicx}
\usepackage{amssymb}
\usepackage{hhline}
\usepackage{array}
\usepackage{booktabs}
\usepackage{tikz}
\usepackage{url}
\usepackage{hyperref}
\usepackage[normalem]{ulem}

\DeclareMathOperator{\synth}{synth}
\DeclareMathOperator{\Uniform}{Uniform}
\def\E{{\mathbb E}}
\newif\ifdraft
\drafttrue
\draftfalse

\definecolor{dkgreen}{RGB}{0,179,36}
\definecolor{dkred}{RGB}{240,0,0}
\definecolor{dkblue}{RGB}{0,70,160}
\definecolor{dkorange}{RGB}{230,100,20}
\definecolor{dkbrown}{RGB}{170,50,0}
\definecolor{chestnut}{rgb}{0.8, 0.36, 0.36}
\definecolor{pink}{RGB}{255,0,247}
\definecolor{amber}{rgb}{1.0, 0.75, 0.0}
\definecolor{amethyst}{rgb}{0.6, 0.4, 0.8}

\newcommand{\kl}[1]{\ifdraft \textcolor{red}{[{\it KL: #1}]} \fi}
\newcommand{\cmc}[1]{\ifdraft \textcolor{dkbrown}{[{\it CMC: #1}]} \fi}
\newcommand{\jc}[1]{\ifdraft \textcolor{dkblue}{[{\it JC: #1}]} \fi}

\ifdraft
\newcommand{\kledit}[1]{\textcolor{blue}{{#1}}}
\newcommand{\kleditcr}[1]{\textcolor{red}{{#1}}}
\newcommand{\cmcedit}[1]{\textcolor{olive}{{#1}}}
\newcommand{\jcedit}[1]{\textcolor{dkgreen}{{#1}}}
\newcommand{\jceditar}[1]{\textcolor{dkgreen}{{#1}}}
\newcommand{\klremove}[1]{\textcolor{blue}{{\sout{#1}}}}
\newcommand{\klremovecr}[1]{\textcolor{red}{{\sout{#1}}}}
\newcommand{\cmcremove}[1]{\textcolor{olive}{{\sout{#1}}}}
\newcommand{\jcremove}[1]{\textcolor{dkgreen}{{\sout{#1}}}}
\newcommand{\todo}[1]{\textcolor{dkred}{[{\it ToDo: #1}]}}
\newcommand{\note}[1]{\textcolor{dkred}{[{\it Note: #1}]}}
\newcommand{\mm}[1]{\textcolor{dkred}{[{\it mark: #1}]}}
\else
\newcommand{\kledit}[1]{\textcolor{black}{{#1}}}
\newcommand{\kleditcr}[1]{\textcolor{black}{{#1}}}
\newcommand{\cmcedit}[1]{\textcolor{black}{{#1}}}
\newcommand{\jcedit}[1]{\textcolor{black}{{#1}}}
\newcommand{\jceditar}[1]{\textcolor{black}{{#1}}}
\newcommand{\klremove}[1]{{}}
\newcommand{\klremovecr}[1]{{}}
\newcommand{\cmcremove}[1]{{}}
\newcommand{\jcremove}[1]{{}}
\newcommand{\todo}[1]{}
\newcommand{\note}[1]{}
\newcommand{\mm}[1]{{}}
\fi

\usepackage[subtle]{savetrees}

\title{AV2Wav: Diffusion-based re-synthesis from continuous self-supervised features for audio-visual speech enhancement}

\name{Ju-Chieh Chou, Chung-Ming Chien, Karen Livescu}
\address{Toyota Technological Institute at Chicago}

\begin{document}
%
\maketitle


%
\begin{abstract}
Speech enhancement systems are typically trained using pairs of clean and noisy speech. \klremove{However, obtaining clean audio-visual speech can be difficult, as many existing audio-visual speech datasets are collected from real-world environments and inherently contain natural background noise, which limits the development of audio-visual speech enhancement systems.}\kledit{In}\jcremove{ the case of}\jc{to be in the word limit}\kledit{ audio-visual speech enhancement (AVSE), there is not \jcremove{nearly }as much ground-truth clean data available; most audio-visual datasets are collected in real-world environments with background noise and reverberation, hampering the development of AVSE.} 
In this work, we introduce AV2Wav, a resynthesis-based audio-visual speech enhancement approach that can generate clean speech despite the challenges of real-world training data.\cmc{What's the most intriguing part of this work? I believe there is something being worth mentioning other than ``simple''.}\jc{I think it will be by filtering and noise-robust training, it can possibly generate better audio quality than the original target. But it is hard to describe this in a few words. We seem to also emphasize it quite a lot.}
We \klremove{first}obtain a subset of nearly clean speech from \klremove{the}\kledit{an} audio-visual corpus using a neural quality estimator\klremove{. We}\kledit{, and} then train a diffusion model \kledit{on this subset} to generate waveform\kledit{s} condition\klremove{ing}\kledit{ed} on \kledit{continuous} \jcremove{audio-visual self-supervised }speech representations \klremove{, i.e.}\kledit{from} \jcremove{noise-robust }AV-HuBERT \jcedit{with noise-robust training}\kledit{.}\klremove{, on the nearly clean subset.}
\kledit{ We use continuous rather than discrete\jcremove{AV-HuBERT} representations\jcremove{ in order} to retain prosody and speaker information.} 
\jcremove{ We find that }With this vocoding task alone, the model can perform speech enhancement better than a masking-based baseline.\jcremove{trained using clean/noisy pairs}\jcremove{due to the noise robustness of AV-HuBERT representations}
\jcremove{To further improve the performance and robustness, }
We further fine-tune the model on clean/noisy utterance pairs \jcedit{to improve the performance.}\jc{should we say performance}\cmc{maybe we can mention that how much paired data we use in the abstract if it's an impressive point}\jc{130 hrs, probably not impressive} \klremove{Our result shows that o}\kledit{O}ur approach outperforms \klremove{the}\kledit{a} masking-based baseline in \kledit{terms of both} automatic metrics and \klremove{the}\kledit{a} human listening test\cmc{can we say our approach is comparable with SoTA models? It's not clear how good this baseline is if the readers only quickly pass through the abstract.}\jcremove{ and is capable of generating high-quality and natural-sounding speech}\kledit{ and is close in quality to the target speech in the listening test}. 
\footnote{Audio samples can be found at \href{https://home.ttic.edu/~jcchou/demo/avse/avse_demo.html}{\url{https://home.ttic.edu/~jcchou/demo/avse/avse_demo.html}}. }

\end{abstract}
\vspace{-.1in}
\section{Introduction}
\vspace{-.1in}
\begin{figure}[h]
    \centering
    \includegraphics[width=0.95\linewidth]{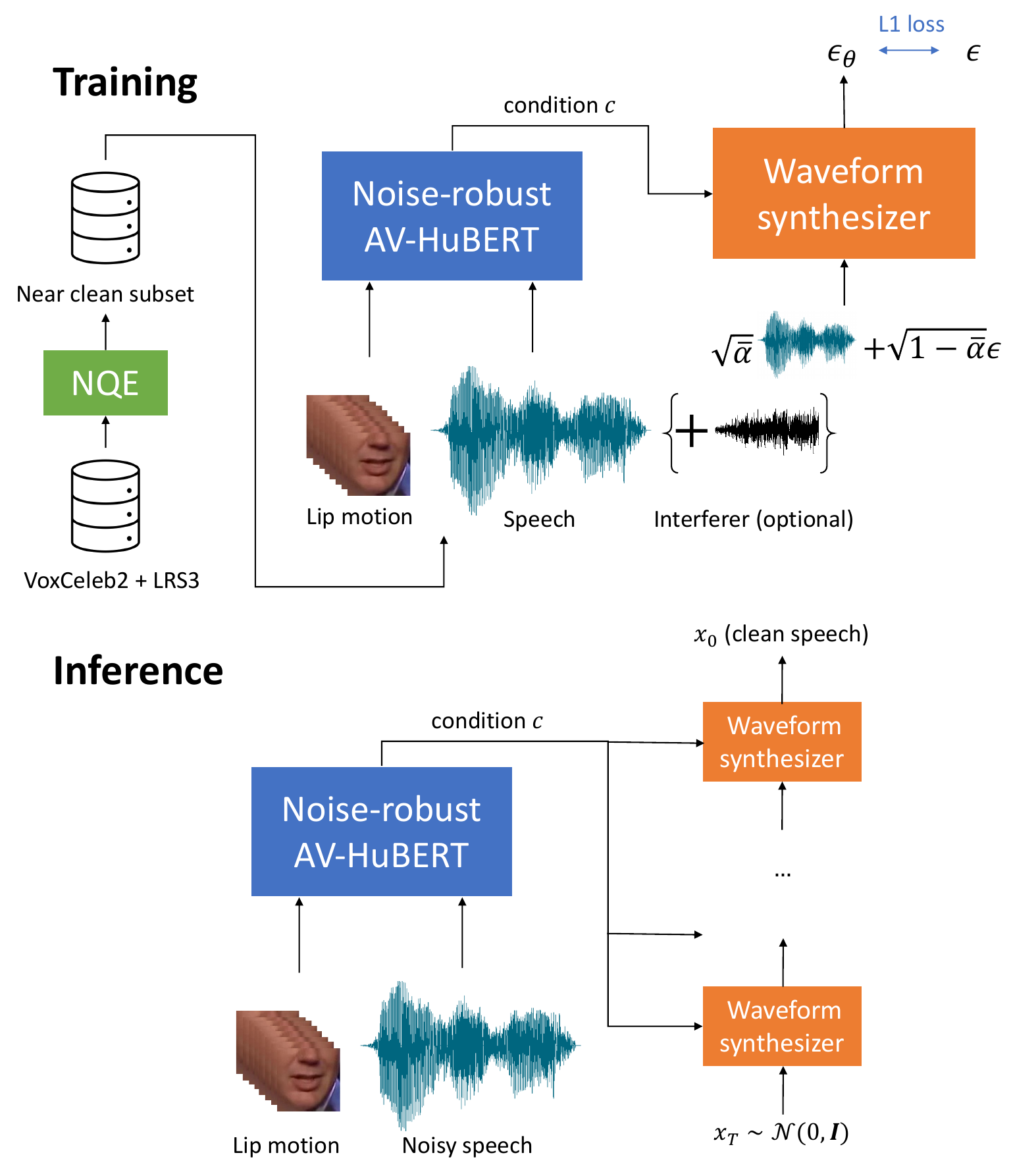}
    \caption{Overview of our \klremove{model.}\kledit{approach.} 
    We obtain a nearly clean subset of the AV \klremovecr{dataset}\kleditcr{training set (VoxCeleb2 + LRS3)} using a neural quality estimator\kleditcr{ (NQE)} and use noise-robust AV-HuBERT to encode the AV speech\klremove{ representations}. These representations are used as condition\kleditcr{ing input}\klremove{s to train}\kledit{ to} a diffusion-based waveform synthesizer\klremove{ to re-synthesize the original waveform}.\cmc{It's not clear what the role of the Interferer is. Also we should move the ``plus'' symbol to the space between $x_0$ and the black waveform.}\jc{No, the interferer is added to the input of the noise robust avhubert. } 
    \jc{Change the size of interferer}}
    \label{fig:model}
\end{figure}

\cmc{An important question: I assume AV2Wav refers to the diffusion model but according to Fig. \ref{fig:model}, but later in the expperiment section it looks like the concept covers everything in this paper, which make many many things confusing.}\jc{I see. We should make the distinction.}
\jcedit{Speech enhancement aims to improve the audio quality and intelligibility of noisy speech. Audio-visual speech enhancement (AVSE)} \klremove{is to }use\kledit{s} visual cues, \klremove{e.g., talking face}\kledit{specifically video of the speaker}\klremove{'s face}, to \jcremove{recover information that is lost in noisy speech signals and}\jcedit{improve the performance of speech enhancement.} 
\cmcedit{Visual cues can provide}\jcedit{ auxiliary information, such as the \klremove{lip motion and the} place of articulation}\jcremove{ both phonetic fine-grained information e.g., the place of articulation}\cmcedit{, which is especially useful when the}\jcremove{speech signals are too noisy to distinguish.} signal-to-noise ratio\klremove{(SNR)} is low. 
 
 \jcedit{Conventionally, }\cmcedit{audio-visual speech enhancement}\jcedit{ is} formulated as a mask regression problem.\cmc{Existing solutions include mask-based methods and re-synthesis-based methods.} Given a noisy utterance and its corresponding \klremove{talking face, }\kledit{video, masking-based models attempt to recover the \klremove{original}clean speech by multiplying}\cmcremove{such that} the noisy signal with a learned mask\cmcremove{can recover the original signal}~\cite{hou2018audio,chern2023audio,afouras2018conversation,gao2021visualvoice}. 
 \jcremove{However, it can be difficult to recover the original signal with one mask operation.}\jcedit{However, some signals are difficult or even impossible to reconstruct via masking. Masking operations tend to allow noise to bleed through, and they cannot effectively address unrecoverable distortion, such as frame dropping.}
 
\klremove{Recently works}\kledit{Some work} \klremove{propose}\kledit{has proposed} to formulate SE and AVSE as a synthesis or re-synthesis problem~\cite{yang2022audio,hsu2023revise,pascual2017segan,richter2023speech,serra2022universal}. \klremove{They}\jceditar{Re-synthesis based approaches} learn discrete audio-visual representations \cmcedit{from} \cmcremove{between}clean speech \cmcedit{and train models to generate} \jcedit{the discrete representations}\jcremove{ derived}\jcedit{ of the clean speech given the corresponding noisy speech.}\cmcremove{as "content"}
 \cmcremove{and then use a another}\cmcedit{ An} off-the-shelf vocoder \cmcedit{trained on clean speech is then used} to \cmcremove{recover the original}\cmcedit{produce clean speech} signals\jcremove{given the discrete representations}.
 This formulation can better handle unrecoverable distortion and synthesize speech with better audio quality. \kledit{However, such discrete representations often lose much of the speaker and prosody information~\cite{polyak21_interspeech}}.

\klremove{On the other hand,}\kledit{Another challenge in AVSE is the suboptimal audio quality of audio-visual (AV) datasets.}\jcremove{Unlike studio-recorded speech-only datasets, }\klremove{hinders the development of audio-visual speech enhancement. \mm{}\jcedit{AV datasets\kledit{,} in contrast to studio-recorded speech-only datasets,}} \klremovecr{In contrast to studio-recorded speech-only datasets, AV datasets are \jcedit{either} collected \kledit{"in the wild" with varying recording environments and natural noise~\cite{chung2018voxceleb2,afouras2018lrs3}}\jcedit{ or limited in the amount of data~\cite{cooke2006audio}.}}\kleditcr{In contrast to studio-recorded speech-only datasets, clean AV datasets (e.g.,~\cite{cooke2006audio}) are much smaller, so many researchers (including ourselves) resort to more plentiful but less clean AV data collected ``in the wild"~\cite{chung2018voxceleb2,afouras2018lrs3}.} \klremove{The recording environments vary and may contain natural noise.}\jcremove{Using these suboptimal data as clean data to train enhancement models leads to suboptimal results.}

In this work, we propose AV2Wav, \kledit{a re-synthesis-based approach to AVSE that addresses the challenges of noisy training data and lossy discrete representations (see Fig~\ref{fig:model}).}\klremove{as shown in Fig~\ref{fig:model}} Instead of \klremove{using} discrete representations,\klremove{which can lose speaker and prosodic information,} we use continuous features from a pre-trained noise-robust AV-HuBERT~\cite{shi2021learning}\kledit{, a self-supervised audio-visual speech model,} \klremove{trained by self-supervision,}\cmcremove{representations as a condition} \cmcedit{to condition \klremove{our}\kledit{a} diffusion-based waveform synthesizer~\cite{chen2020wavegrad}. }\cmcedit{The noise-robust training enables AV-HuBERT to generate similar representations given clean or \jcedit{mixed}\jcremove{interfered}
(\kledit{containing}\klremove{i.e.,} noise or a competing speaker) speech. }\jcremove{ As the noise-robust AV-HuBERT has been trained to learn noise-invariant features, we use it as a noise-removal model to remove noise.}\jcremove{AV-HuBERT is a self-supervised model for audio-visual speech, and has been shown to }\jcremove{be useful in learning phonetic representations~\cite{pasad2023comparative}}\jcremove{ and waveform synthesis.}\jcremove{The noise-robust training enables AV-HuBERT to generate similar representations given clean or interfered (i.e., noise or a competing speaker) speech.}\jcremove{ With the AV-HuBERT learning noise-robust representations, the diffusion-based waveform synthesizer can thus generalize to noisy input utterances. }

\klremovecr{\mm{}\jcedit{Self-supervised models have been used for speech enhancement~\cite{huang2022investigating,irvin2023self,hung2022boosting}. }\kledit{Several AVSE approaches have also used AV-HuBERT, but they have either done so for mask prediction~\cite{chern2023audio}, for synthesis of a single speaker's voice~\cite{hsu2023revise}, or with access to transcribed speech for fine-tuning~\cite{richter2023audio}.}}

\jcedit{In addition\klremove{ to the noise-robustness of AV-HuBERT}, we train\klremove{ed} the \klremove{waveform}synthesizer on a nearly clean subset of \klremove{the}\kledit{an} audio-visual \klremove{training} dataset filtered by a neural quality estimator (NQE) to exclude \kledit{low-quality} utterances\klremove{with strong background noise and reverberations to make sure the generated speech is sufficiently clean}.} \kledit{Finally, w}\klremove{W}e further fine-tune\klremove{d} the \jcremove{waveform synthesizer}\jcedit{model} on clean/noisy utterance pairs and studio-recorded clean speech\klremove{ to improve the audio quality and intelligibility}.
\jcremove{Speech enhancement is mainly formulated as a regression problem. }
\jcremove{Traditionally, speech enhancement models are trained with paired clean and noisy speech -- which can be obtained by adding artificial noises to studio-quality speech data.}\cmcremove{Given clean and noisy speech pairs, models are trained to generate the original clean speech given noisy speech.}\jcremove{ However, in audio-visual speech enhancement, where the talking head is provided, datasets are mostly collected in the wild}\jcremove{, for example, interviews, TED talks, and so on}
\jcremove{The recording environments vary and may contain natural noise. Using these suboptimal data as clean data to train enhancement models leads to suboptimal results. }
\jcremove{As a result, we apply a neural quality estimator (NQE) to select a subset of audio-visual data with relatively high speech quality for the training of AV2Wav. We can therefore}\jcremove{get rid of the requirement of paired clean/noisy data and also furtherensure that the AV2Wav only learns to generate clean speech}\jcremove{ even though noisy inputs may be given during the inference time}

\jcremove{AV-HuBERT, an audio-visual self-supervised  model, is trained to predict cluster assignment in the masked region given speech features and lip motion sequences. }\jcremove{This objective has been shown to learn phonetic information~\cite{pasad2023comparative} and benefits downstream tasks such as audio-visual speech recognition~\cite{shi22_interspeech}. AV-HuBERT also utilizes noise-robust training along with mask prediction by predicting the original cluster assignment but given speech signals with an interferer, i.e., noise or a competing speaker~\cite{shi22_interspeech}.}\jcremove{Because the lip motion sequence is given, the interferer can be either noise or another speaker's speech. The model learns noise-invariant representations with noise-robust training. }
\jcremove{We utilize the compact representations learned by AV-HuBERT for audio-visual speech enhancement (AVSE). Conditional diffusion models are trained to generate a waveform conditioned on it AV-HuBERT representation}
\jcremove{In contrast to prior works focusing on training models to remove noise from speech, we formulate the problem as a re-synthesis problem and focus on the vocoding side rather than improving the noise-removal model.}
\jcremove{We use the pre-trained frozen AV-HuBERT model as our noise-removal model to encode noise-invariant representations and then synthesize the waveform from the representations as in Fig~\ref{fig:model}. With the AV-HuBERT learning noise-robust representations, the diffusion model trained with this vocoding task can be generalized to noisy utterances.}\jcremove{ Furthermore, since we only train our diffusion model on a relatively clean subset filtered by neural quality estimation (NQE), our model learns to generate only clean speech.} 
\cmcremove{Previous works on re-synthesis utilize discrete representations in the middle as the "content" of the utterance and then recover the waveform using another vocoder~\cite{yang2022audio}. The discretization step leads to losses of the speaker and prosody information \cmc{already mentioned in previous paragraphs}, which are also important in human interaction. In contrast to discrete representations, we use continuous representations to recover the original waveform to keep most of the information in the original utterance. Furthermore, using continuous representations simplifies the training pipeline and reduces the effort in developing AVSE system.}
\jcremove{Recent work uses fine-tuned AV-HuBERT on video speech recognition (VSR) as video conditioning, and noisy speech as audio condition, to synthesize clean speech using a diffusion model~\cite{richter2023audio}. Our work, on the other hand, uses pre-trained AV-HuBERT, without having access to transcriptions. Also, we resynthesize waveform from AV-HuBERT representations so that the model can be trained on this vocoding task without using clean/noisy speech pairs. }
\jcremove{ We also study the effect of fine-tuning on clean and synthesized noisy speech pairs and studio-recorded clean speech. We fine-tune the diffusion model on the AVSE challenge dataset to understand the effect of fine-tuning on clean and noisy utterance pairs. On the other hand, the near-clean subset we filtered from the audio-visual dataset still contains some noise and reverberation in the speech. We thus also fine-tune the diffusion model on studio-recorded clean speech to see if the audio quality can be further improved. }

The contributions of this work include: (i)\klremove{we present} the \kledit{AV2Wav} \klremove{a simple}framework for \kledit{re-synthesis based} AVSE\klremove{by synthesizing waveform conditioning} \kledit{conditioned} on\klremove{the} noise-robust AV-HuBERT representations; (ii) \klremove{we show}\kledit{a demonstration} that \cmcedit{an} NQE can be used \kledit{for training data selection}\cmcremove{to filter a near-clean subset of speech} to improve \kledit{AVSE performance}\klremove{the quality of the generated speech}; \kledit{and} (iii) \klremove{we}\kledit{a} study \kledit{on} the effect of fine-tuning \jcedit{diffusion-based waveform} \kledit{synthesis}\klremove{further} on clean/noisy data and studio-recorded data.  \kleditcr{The resulting enhancement model outperforms a baseline masking-based approach, and comes close in quality to the target speech in a listening test.}
\klremove{\kledit{\bf Audio-visual speech enhancement}
\klremove{Previous works}\kledit{Several previous approaches for AVSE} have used AV-HuBERT\klremove{ in AVSE}~\cite{hsu2023revise,chern2023audio,richter2023audio}. \kl{try not to start a sentence with a citation.  If you want to do so, then list the authors, e.g. "Chern {\it et al.}~\cite{chern2023audio}"}
\jcedit{Chern {\it et al.}}~\cite{chern2023audio} use AV-HuBERT representations along with a neural regressor to predict a mask to recover \kledit{the} clean speech signal\jcedit{s}. \jcedit{Hsu {\it et al.}}~\cite{hsu2023revise} fine-tune \klremove{the} AV-HuBERT to predict discrete tokens derived from clean speech and use a vocoder trained on a high-quality, single-speaker dataset to recover \klremove{the}\kledit{a} waveform from the discrete tokens. \klremove{Although it can synthesize speech with a sample rate (24k) higher than that of the original speech (16k) }\cmc{do we need to mention its sample rate here?}\kl{I removed it -- hope that's OK}\kledit{However}, it loses \klremove{the} speaker and prosodic information. \jc{As the sentence has said masking and re-synthesis, do we need to explicitly emphasize it?}\kl{I don't quite understand JC's question}\jc{I meant do we need to categorize it (masking and re-synth), even if it has been said in the sentence.}
\kledit{Finally, Richter {\it et al.}~\cite{richter2023audio} use}\klremove{Recent work uses fine-tuned} AV-HuBERT \kledit{fine-tuned} on video speech recognition (VSR) as video conditioning, and noisy speech as audio condition\kledit{ing}, to synthesize clean speech using a diffusion model\klremove{~\cite{richter2023audio}}. Our work, on the other hand, uses \kledit{a frozen} pre-trained AV-HuBERT\klremove{, without having}\kledit{ and requires no } access to transcriptions. \jcedit{We use the noise-robust AV-HuBERT as our noise removal model while \klremove{they only use}\kledit{\cite{richter2023audio} uses} AV-HuBERT \kledit{only} as a video encoder, and \klremove{rely}\kledit{relies} on the diffusion model\klremove{s} to remove noise.} \klremove{We also proposed to use NQE for data selection. }\kl{removed sentence for space}\jcremove{Also, we re-synthesize \kledit{the} waveform from AV-HuBERT representations }\jcremove{so that the model can be trained on this vocoding task without using clean/noisy speech pairs.}  \kl{it is not clear what is the difference you are pointing out} \cmc{I would suggest categorize related works into mask-based and re-synthesis-based methods too.}\jc{Change the second point to NQE} }

\klremove{\kledit{{\bf Diffusion models for speech generation}}
Diffusion models \klremove{are}\kledit{have been} \jcremove{\cmcedit{commonly}} used in \kledit{multiple} speech generation \cmcedit{tasks, such as} \cmcremove{in} \klremove{speech} enhancement~\cite{lu2022conditional,serra2022universal,lu2021study}, text-to-speech\kledit{ synthesis } (TTS)~\cite{huang2022fastdiff,chen2020wavegrad}, and vocod\klremove{er}\kledit{ing}~\cite{kong2020diffwave, chen2020wavegrad}. \jcremove{By using generative models instead of regression objectives, }}
\klremove{\jcedit{Generative model objective\kledit{s} }can generate \jcremove{a more }complex distribution\kledit{s} and handle unrecoverable distortion~\cite{hsu2023revise,serra2022universal}. Diffusion models also have the advantage of stable training, which makes it easier to fine-tune the model on the desired target data domain. \kl{say why you are mentioning this related work.  are you borrowing ideas from it?  is it to contrast with your work?}\jc{To give a bit of context on what is the current progress. The stable training bit is one of the reason to use diffusion models. }\jc{this paragraph seems less important}}

\vspace{-.1in}
\section{Method}
\vspace{-.05in}

\subsection{\kledit{Background: } AV-HuBERT}\label{sec:avhubert}
\mm{}\kleditcr{Self-supervised models are increasingly being used for speech enhancement~\cite{irvin2023self,hung2022boosting,huang2022investigating}. For AVSE, several approaches have used AV-HuBERT.  However, unlike our work, this prior work has used AV-HuBERT either for mask prediction~\cite{chern2023audio}, for synthesis of a single speaker's voice~\cite{hsu2023revise}, or with access to transcribed speech for fine-tuning~\cite{richter2023audio}.}

AV-HuBERT~\cite{shi2021learning,shi22_interspeech} is a self-supervised model trained \klremovecr{with mask\kledit{ed} prediction given}\kleditcr{on} speech and lip motion \kledit{video} sequence\kledit{s}.  
\klremove{After pre-training\jcedit{ on predicting the clustering assignment for the masked region}, for a speech}\kledit{The model is \klremove{pre-}trained to predict a discretized label\jcremove{ (a cluster assignment)} for a masked region of the audio} feature sequence $Y^a_{1:L} \in \mathbb{R}^{F_s \times L}$ and \klremove{lip motion}\kledit{video} sequence $Y^v_{1:L}  \in \mathbb{R}^{F_l \times L}$ with \klremove{sequence length} $L$ \kleditcr{frames and feature dimensionalities $F_s$, $F_l$}\klremove{, the}\kledit{. The resulting} \klremove{AV-HuBERT }model $\mathcal{M}$ \klremove{fuse the two sequences to encode}\kledit{produces} audio-visual representations \klremove{as} 
\begin{equation}
f^{av}_{1:L}=\mathcal{M}(Y^a_{1:L}, Y^v_{1:L})  
\label{eq:f_av}
\end{equation}
\klremovecr{After pre-trained by mask prediction. }AV-HuBERT \klremovecr{applies}\kleditcr{uses} modality dropout during training~\cite{neverova2015moddrop}, i.e.\kleditcr{ it} drops one of the modalities with \kleditcr{some} probability, to learn modality-agnostic representations\klremovecr{,}
\begin{equation}
\begin{split}
    f^a_{1:L}=\mathcal{M}(Y^a_{1:L}, \mathbf{0}), \\
    f^v_{1:L}=\mathcal{M}(\mathbf{0}, Y^v_{1:L}),
\end{split}
\label{eq:f_a_v}
\end{equation}
Some versions of AV-HUBERT \klremove{are trained with}\kledit{use} noise-robust training~\cite{shi22_interspeech}\klremovecr{.} 
\kledit{, where an interferer (noise or competing speech) is added while the model must still predict cluster assignments learned from clean speech.} 
\kledit{In this case the model outputs the \klremove{audio-visual} representation}
\begin{equation}
    f^{avn}_{1:L}=\mathcal{M}(\synth(Y^a_{1:L}, Y^n_{1:L}), Y^v_{1:L}), 
\label{eq:f_avn}
\end{equation}
\kledit{where} $\synth(\cdot)$ is a function \klremove{to synthesize}\kledit{that synthesizes} noisy speech given noise $Y^n_{1:L}$ and speech $Y^a_{1:L}$.\kl{do we need $\synth$, or is it equivalent to $Y^a_{1:L} + Y^n_{1:L}$ ?}\jc{it could be mixed with some SNR. Let me check.}\jc{there might be asynth() in the original paper (code), not clear}\cmc{Do we also need to say the model is trained with similar classification labels no matter $f^{??}$ is used?}
\jcedit{Noise-robust AV-HuBERT is trained to predict the same clustering assignment given $f^{a}, f^{v}, f^{av}$ and $f^{avn}$\kledit{, in order} to learn modality\kledit{-} and noise-invariant features.} As \klremove{the}AV-HuBERT \klremove{model}\cmcedit{already learns to remove noise \klremove{information}through the}\cmcremove{is already trained to learn} noise-invariant \cmcedit{training}\cmcremove{representations}, \kledit{it is a natural choice as a conditioning input to our AVSE model.} \klremove{we utilize the noise-robust AV-HuBERT representations as a condition \cmcedit{in our audio-visual speech enhancement network} \cmcremove{to perform speech re-synthesis}. }

\vspace{-.1in}
\subsection{Diffusion waveform synthesizer}\label{sec:diffusion}
Our \jcedit{diffusion-based waveform synthesizer}\jcremove{AV2Wav model} is \jcremove{a diffusion model }based on WaveGrad~\cite{chen2020wavegrad}. \kleditcr{We summarize the formulation here; for details see~\cite{chen2020wavegrad}.} \mm{}\klremovecr{\jcedit{We modify the original up-sampling rate \jcedit{of WaveGrad} to generate waveform conditioning on the AV-HuBERT features. }}For speech waveform $x_0 \in \mathbb{R}^{L_w}$ with length $L_w$, 
the diffusion forward process is formulated as a Markov chain to generate $T$ latent variable\kledit{s} $x_1, \dots, x_T$\jcedit{ with the same dimensionality as $x_0$,}
\begin{equation}
    q(x_1, x_2, \dots, x_T|x_0)=\prod_{t=1}^T q(x_t|x_{t-1}),
\end{equation}
\kledit{where} $q(x_t|x_{t-1})$ is a Gaussian distribution\klremove{,}\kledit{:}
\begin{equation}
    q(x_t|x_{t-1})=\mathcal{N}(x_t; \sqrt{1-\beta_t} x_{t-1}, \beta_t \mathbf{I})
\end{equation}
with \kledit{a} pre-defined noise schedule $0 < \beta_1 < \beta_2 \dots < \beta_T < 1$. 
\jcedit{The idea is to gradually add noise to the data distribution, \klremove{so}\kledit{until} $P(x_T)$ is close to a multivariate Gaussian distribution with zero mean and unit variance: $p(x_T) \approx \mathcal{N}(\kledit{x_T; }0, \mathbf{I})$. }We can also directly sample from $q(x_t|x_0)$ by reparameterization,
\begin{equation}
\label{eq:forward}
    q(x_t|x_0)=\mathcal{N}(x_t;\sqrt{\bar{\alpha_t}} x_0, (1-\bar{\alpha_t})\mathbf{I})
\end{equation}
where $\alpha_t=1-\beta_t$ and $\bar{\alpha}_t=\prod_{i=1}^t \alpha_i$. 

The reverse process is parameterized by a neural network $\epsilon_\theta(\cdot)$\kleditcr{, which takes the noised waveform drawn from Eq.~\ref{eq:forward}, the conditioning input (here, the AV-HuBERT features), and a noise level, and outputs a prediction of the added Gaussian noise in Eq.~\ref{eq:forward}.} \klremovecr{to predict the added Gaussian noise given the diffused waveform, conditioning input and noise level. }\jcremove{We use the $\epsilon$-prediction\kledit{ approach~\cite{ho2020denoising} and continuous noise level~\cite{chen2020wavegrad}. }}\jcremove{, namely training a neural network $\epsilon_\theta(\cdot)$ to predict the Gaussian noise $\epsilon$ added to the data $x_0$ given condition $c$ (AV-HuBERT features with dimension $F$ and length $S$) and noise level $\sqrt{\bar{\alpha}}$,}\jcremove{proposed in~\cite{ho2020denoising,chen2020wavegrad}.}\jcremove{ and the continuous noise level\jcremove{conditioning }\klremove{in}\jcremove{of}~\cite{chen2020wavegrad} to predict $\epsilon$\kl{define $\epsilon$}\jc{mentioned in prev sentence ($\epsilon$-prediction)} condition\klremove{ing}\kledit{ed} on the AV-HuBERT feature segment $c \in \mathbb{R}^{F \times S}$\jcedit{ with feature dimension $F$ and length $S$}.
We \jcedit{uniformly }sample a segment\jcedit{ of AV-HuBERT feature frames} with length $S$}\jcremove{\kleditcr{:}} \kleditcr{In training, we}\klremovecr{We} first sample AV-HuBERT features\klremovecr{ as}
\jcremove{ from the AV-HuBERT representations} 
\klremovecr{\begin{equation}
\label{eq:seg}
    l=\Uniform(1, L-S+1)
\end{equation}
\mm{}and sample AV-HuBERT features,}
\begin{equation}
    c=\begin{cases}
        f^{av}_{l:l+S} & \text{with probability } p_{av} \\
        f^{a}_{l:l+S} & \text{with probability } p_{a} \\
        f^{v}_{l:l+S} & \text{with probability } p_{v} \\
        f^{avn}_{l:l+S} & \text{with probability } p_{avn} \\
    \end{cases}
\end{equation}
where \kleditcr{$l=\Uniform(1, L-S+1)$, }$p_{av}+p_a+p_v+p_{avn}=1$\kledit{ and $f^{a}, f^{v}, f^{av}, f^{avn}$ are as defined in Eq.~\ref{eq:f_av}, \ref{eq:f_a_v}, \ref{eq:f_avn}.}

We then sample a continuous noise level $\sqrt{\bar{\alpha}}$\kledit{:}\klremove{ as,}
\begin{equation}
    s \sim \Uniform(\{1 \dots T\}),
\end{equation}
\begin{equation}
    \sqrt{\bar{\alpha}} \sim \Uniform(\sqrt{\bar{\alpha}_{s-1}}, \sqrt{\bar{\alpha}_{s}}),
\end{equation}
\mm{}and minimize
\begin{equation}
    \E_{x_0, c,\sqrt{\bar{\alpha}}}[\Vert \epsilon - \epsilon_\theta(\sqrt{\bar{\alpha}}x_0 + \sqrt{1-\bar{\alpha}} \epsilon,c, \sqrt{\bar{\alpha}})\Vert_1]
\end{equation}
\kledit{where} $x_0$ is the \klremove{corresponding}waveform segment \klremove{of}\kledit{corresponding to} $c$\kleditcr{, as in~\cite{chen2020wavegrad} and $\epsilon \sim \mathcal{N}(0, \textbf{I})$.}\klremovecr{\jcremove{, i.e, $x_0, c \sim p(x_0, c)$}.
As in~\cite{chen2020wavegrad}, we \klremove{train the model with}\kledit{use} L1 loss to stabilize training. }
After training $\epsilon_\theta$, we can sample from \jcedit{$p_\theta(x_0|c)$ by re-parameteriz\kledit{ing} \klremovecr{the }$\epsilon_\theta$\kleditcr{:}\klremovecr{ as,}}  
\begin{equation}
    p_\theta(x_0|c)=p(x_T|c)\prod_{t=1}^{T} p_{\theta}(x_{t-1}|x_t, c)
\end{equation}
\begin{equation}
   p_\theta(x_{t-1}|x_t, c)=\mathcal{N}(x_{t-1};\mu_\theta(x_t, t), \tilde{\beta_t}) 
\end{equation}
where $\mu_\theta(x_t, t)=\frac{1}{\sqrt{\bar{\alpha}_t}}(x_t - \frac{1-\alpha_t}{\sqrt{1-\bar{\alpha}_t}}\epsilon_\theta(x_t, c, \sqrt{\bar{\alpha}_t}))$, \klremovecr{and }$\tilde{\beta}_t=\frac{1-\bar{\alpha}_{t-1}}{1-\bar{\alpha}_{t}} \beta_t$, and $p_(x_T|c)\approx\mathcal{N}(x_T;0, \textbf{I})$. 

\subsection{\klremove{Training d}\kledit{D}ata filtering with a \jcedit{n}eural \jcedit{q}uality \jcedit{e}stimator (NQE)}\label{sec:nqe}
\vspace{-.05in}
\kl{Looks like you switched to lower-case "neural quality estimator" in the subsection heading.  Make sure it is consistent throughout the paper.}
\jcremove{Audio-visual datasets collected in the wild are likely to contain some natural noise. If we train the }\jcremove{generation model with noisy data} \cmcremove{to generate speech with natural noise}\jcremove{, the model will generate a distribution with such noise, which is against the purpose of speech enhancement. }

We propose to use \kledit{a }\jcremove{ neural quality estimator (}NQE to select a relatively clean subset from the \kledit{training} \klremove{data}set\klremove{ to train the AV2Wav model}. \jcedit{Conventional quality metrics (e.g., SI-SDR~\cite{le2019sdr}) require \kledit{a} reference} \kledit{signal, which \kledit{our training data lacks.}\klremove{we don't have for our training data.}\klremove{ to compute.}} NQE \klremove{aims to predict }\kledit{predicts} \klremove{the}\kledit{an} audio quality metric without a reference using a neural network. \klremove{In the case of data selection, we don't have a reference for the audio. Thus, w}\klremovecr{\kledit{W}e use NQE\kledit{---specifically,}}\kleditcr{Specifically, we use the predicted scale-invariant signal-to-distortion ratio (P-SI-SDR) of~\cite{kumar2023torchaudio}, and retain those utterances with P-SI-SDR above some threshold.}\klremovecr{\jcremove{as }in~\cite{kumar2023torchaudio}---as a proxy to select utterances that are sufficiently clean. \mm{}We also use P-SI-SDR as an evaluation metric to measure audio quality.}
\klremove{\klremove{W}\kledit{Specifically, w}e use \klremove{the prediction of the}\kledit{a predicted} scale-invariant signal-to-distortion ratio (SI-SDR)~\cite{le2019sdr}, in~\cite{kumar2023torchaudio} as our quality estimator. }\jcremove{~\cite{kumar2023torchaudio} is a neural quality estimator trained with multi-tasking to predict the reference-based metric without providing reference audio. }
\cmc{give some details about the NQE here; otherwise it does not look like provide any other information than things we've already known in the introduction.}\jc{maybe this can be remove}

\vspace{-.125in}
\section{Experiments}
\vspace{-.1in}
\subsection{Datasets and baseline}
\vspace{-.05in}
\klremove{Our models are trained in two stages. }In the first \kledit{training} stage\kledit{ of our models}, we use the combination of LRS3~\cite{afouras2018lrs3} and \jcedit{an English subset \kledit{(selected using Whisper-large-v2~\cite{radford2023robust})} of VoxCeleb2~\cite{chung2018voxceleb2} (total 1967 hours)}\klremove{ filtered using Whisper-large-v2~\cite{radford2023robust} language identification}. \jcedit{In this stage, we train\klremove{ed} the waveform synthesizer to synthesize waveforms from AV-HuBERT features.}\jc{it will be said in the training section, do we need it here?}\jcremove{As VoxCeleb2~\cite{chung2018voxceleb2} has non-English utterances, we use Whisper-large-v2~\cite{radford2023robust} language identification to filter English utterances, resulting in about 1967 hours of speech. }
In the second stage, we fine-tune the model on noisy/clean \kledit{paired} data from the AVSE challenge~\cite{blanco2023avse}\kledit{, containing 113 hours of speech}. \jceditar{The AV-HuBERT model is frozen unless stated explicitly. }\klremove{The training set contains 113 hours of speech.}Interferers\klremove{(the added audio to distract from the speech)}
include noise and \kledit{speech from a }\jcedit{competing} speaker. The noise sources are sampled as in~\cite{blanco2023avse}\jcremove{\kledit{the} Clarity Challenge~\cite{graetzer2021clarity}, DEMAND~\cite{thiemann2013diverse}, \kledit{and the} DNS Challenge~\cite{reddy2021icassp}}. Competing speakers are sampled from LRS3~\cite{afouras2018lrs3}.

\kledit{For evaluation, w}\klremove{W}e follow\klremove{ed} the recipe provided by the AVSE challenge~\cite{blanco2023avse} to synthesize a test set \kledit{of clean/noisy pairs based on}\klremove{ using the utterances of} the LRS3 test set\klremove{ for our evaluation}. 
We sample\klremove{d} 30 speakers from the LRS3 test set as competing speakers\jcremove{, and the rest as clean speech}. Noise interferers are sampled \jcremove{from }\klremove{among the noise audio files in}\kledit{from the same noise datasets as in~\cite{blanco2023avse} (but excluding the files used in training/dev).} 
\klremove{not used in the AVSE training/dev set.} The SNR is uniformly sampled as in~\cite{blanco2023avse}. 

\kleditcr{As a baseline, we}\klremove{We also} use the \kleditcr{open-source} masking-based baseline \kledit{trained on the AVSE dataset provided }in~\cite{blanco2023avse}\klremove{ as our baseline\jcedit{ to fairly compare to our model as the baseline is trained on the AVSE dataset}}. \kleditcr{For the case of overlapping speakers, we also compare to VisualVoice~\cite{gao2021visualvoice}, the most competitive publicly available audio-visual speaker separation model.  Finally, as a topline, we compare to the target speech.\footnote{Other prior work we are aware of is on either removal of competing speech or removal of noise, whereas our setting combines both; and some models are trained and evaluated on less diverse data, so are difficult to compare with.}}

\vspace{-.15in}
\subsection{Architecture and training}
\vspace{-.05in}
We use the WaveGrad~\cite{chen2020wavegrad} architecture, but adjust the upsampling rate \kledit{sequence} to (5,4,4,2,2,2), resulting in a total \kledit{upsampling rate} of 640, \klremovecr{which can}\kleditcr{to} convert the 25Hz AV-HuBERT \jcedit{features} to a \kleditcr{16kHz} waveform. \jcremove{Our preliminary study shows that adding an extra layer to the down-sampling and up-sampling blocks improves the result, so we used it in all experiments.}We use the \jcedit{features} from the last layer of noise-robust AV-HuBERT-large\kledit{ (specifically the model checkpoint ``Noise-Augmented AV-HuBERT Large")}~\cite{shi22_interspeech,shi2021learning}. \kledit{In training, w}\klremove{W}e uniformly sample \kleditcr{$S=24$} frames from \kledit{the} AV-HuBERT \kledit{features of each utterance}\klremove{during training}
and apply layer normalization to them~\cite{ba2016layer}. \jcremove{We also normalize the waveform by dividing the maximum absolute value through the entire utterance.}
In the first stage, we train\klremove{ed} \klremove{the }AV2Wav \klremove{model }with $(p_{av}, p_{a}, p_v, p_{avn})=(1/3, 1/3, 1/3, 0)$ on the filtered dataset (LRS3 + VoxCeleb2) without adding interferers for 1M steps. We use the Adam optimizer~\cite{kingma2014adam} with a learning rate of $0.0001$ and a cosine learning rate schedule\klremovecr{r} for 10k warm-up steps using a batch size of $32$. In the second stage of training, we fine-tune the model on audio-visual clean/noisy speech \kledit{pairs} with $(p_{av}, p_{a}, p_v, p_{avn})=(0, 0, 0, 1)$ for 500k steps. To understand the effect of fine-tuning, we also fine-tune AV2Wav on VCTK~\cite{veaux2017cstr}, which is a studio-recorded corpus, with $(p_{av}, p_{a}, p_v, p_{avn})=(0, 1, 0, 0)$\jcremove{ to see if we can improve the audio quality by fine-tuning on a clean audio-only dataset}.\jcremove{In the second stage, we also optionally unfreeze the AV-HuBERT model.}

\vspace{-.1in}
\subsection{Evaluation}
\vspace{-.075in}
Signal-level metrics are not ideal for generative models \kledit{because perceptually-similar generated and reference speech may be dissimilar on the signal level.}\klremove{due to the potential dissimilarity between generated and reference speech\jcedit{ \klremove{signals while being}\kledit{that are} perceptually similar. }}  
In addition to objective metrics\kleditcr{---P-SI-SDR~\cite{kumar2023torchaudio} and word error rate (WER)\footnote{We use Whisper-small-en~\cite{radford2023robust} as the ASR model.} as a proxy for intelligibility---}we use \kledit{subjective} human-rated comparison mean opinion scores (CMOS) on a scale of +3 (much better), +2 (better), +1
(slightly better), 0 (about the same), -1 (slightly worse), -
2 (worse), and -3 (much worse)\klremove{for subjective experiments } as in~\cite{loizou2011speech}. \jcedit{We sample 20 pairs for each system and collect at least 8 ratings for each utterance pair. To help listeners better distinguish the quality, we only use utterances longer than 4 seconds and provide the transcription. We use the same instruction\kledit{s} as in~\cite{serra2022universal}\klremove{ to guide our listeners}. Listeners are \kleditcr{proficient (not necessarily native) English} speakers.} 

\vspace{-.15in}
\subsection{Results}
We show the objective evaluation in Table~\ref{tab:mixed}, subjective evaluation in Table~\ref{tab:cmos}, and WER analysis for multiple SNR ranges and interferer types in Table~\ref{tab:error}. 

\vspace{-.05in}
\begin{table}[htbp]
\begin{center}
\caption{\klremove{The Subjective}\kledit{Objective} evaluation \klremove{for }\kledit{in terms of }WER \kledit{(\%)} and \klremovecr{predicted }SI-SDR (P-SI-SDR) (dB) \klremove{by NQE}. 
\textbf{Target re-synthesis} \klremove{part is re-synthesize}\kledit{refers to re-synthesis of} the target \kledit{(clean)} speech using AV2Wav. The \klremove{rest}\kledit{remaining} parts (\textbf{Mixed speech}, \textbf{After fine-tuning}, \textbf{Fast inference}) take \klremove{the}mixed speech as input and synthesize the \kledit{predicted clean} speech (performing AVSE). \kledit{\textbf{Mixed speech} refers to the first stage of AV2Wav training.} \textbf{After fine-tuning} \kledit{refers to} further \kledit{fine-tuning} the \klremove{waveform} synthesizer on \klremove{the }AVSE\jcedit{,} \klremove{or }VCTK\klremove{ dataset}. The \klremove{name is organized}\kledit{model name is given} as AV2Wav-\{filter \kleditcr{criterion}\}-\{fine-tuned dataset\}. \textbf{Fast inference} compares fast inference approaches\kledit{, using}\klremove{. We use} \textit{AV2Wav-23-long-avse} (line \textbf{13}).
\cmc{I would suggest breaking this table into multiple ones and to help the reader focus on the information they need to know when reading a certain experiment. For example, ``after fine-tuning'' and ``fast inference'' can be two independent tables.}\cmc{it's not clear that ``mixed speech'' are the major AVSE results.}\cmc{the target re-synthesis numbers are not explained}\jc{explained in 4.4.4} 
}
\label{tab:mixed}
\begin{tabular}{lrr}
\toprule
& WER $\downarrow$        & P-SI-SDR $\uparrow$            \\ \midrule
              \multicolumn{3}{c}{\textbf{Target re-synthesis}}   \\ \hline
\textbf{(1)} Target  &   6.72         &    21.56                    \\ \hline
\textbf{(2)} AV2Wav-23     &      3.15      &      21.36                 \\ \hline
\textbf{(3)} AV2Wav-23-long     &      \textbf{2.71}      &      \textbf{22.24}                 \\ \hline
\textbf{(4)} AV2Wav-25     &     3.00       &          21.76         \\ \hline
\textbf{(5)} AV2Wav-random &     2.94       &      19.79                 \\ \hline
\multicolumn{3}{c}{\textbf{Mixed speech }}   \\ \hline
\textbf{(6)} Mixed \kledit{input}  &   48.45         &    0.12               \\ \hline
\textbf{(7)} Baseline~\cite{blanco2023avse}  &   26.40         &    13.79                  \\ \midrule
\textbf{(8)} AV2Wav-23     &      18.17      &      19.41                 \\ \hline
\textbf{(9)} AV2Wav-23-long     &      \textbf{17.20}      &      \textbf{20.09}                 \\ \hline
\textbf{(10)} AV2Wav-25     &     17.36       &          20.07          \\ \hline
\textbf{(11)} AV2Wav-random &     19.69       &      14.57                  \\ \midrule
\multicolumn{3}{c}{\textbf{After fine-tuning}}   \\ \hline
\textbf{(12)} AV2Wav-23-avse     &      15.85      &      20.65            \\ \hline
\textbf{(13)} AV2Wav-23-long-avse     &      16.76      &      21.21               \\ \hline
\textbf{(14)} AV2Wav-23-long-avse     &      \textbf{12.77}      &      21.23               \\ 
 (fine-tune AV-HuBERT) & & \\ \hline
\textbf{(15)} AV2Wav-23-vctk     &      16.71      &      \textbf{21.79}                  \\ \hline
\multicolumn{3}{c}{\textbf{Fast inference}}   \\ \hline
\textbf{(16)} AV2Wav-cont-100  &   17.46         &    \textbf{18.43}               \\ \hline
\textbf{(17)} AV2Wav-ddim-100  &   16.21         &    18.17                  \\ \hline
\textbf{(18)} AV2Wav-ddim-50      &      \textbf{15.73}      &      18.15                 \\ \hline
\textbf{(19)} AV2Wav-ddim-25      &      16.41      &      17.58                 \\ \bottomrule
\end{tabular}
\end{center}
\end{table}
\begin{table}[h]
\begin{center}
\caption{\kledit{Comparison mean opinion scores (CMOS) for several model comparisons.  A positive CMOS indicates that the "Tested" model is better than the "Other" model.  The "re-syn" model simply re-synthesizes the target (clean) signal.}
}
\label{tab:cmos}
\begin{tabular}{llr}
\toprule
       \multicolumn{1}{c}{\jcremove{Approach }\jcedit{Tested}}
       & \multicolumn{1}{c}{Other}
       & \multicolumn{1}{c}{CMOS}              \\ \midrule
      AV2Wav-23-long-avse        & Baseline~\cite{blanco2023avse}          &    2.22 $\pm$ 0.16                    \\ \hline
      AV2Wav-23-long-avse        & AV2Wav-23-long          &    0.21 $\pm$ 0.18                    \\ \hline
        AV2Wav-23-long-avse        & Target          &    -0.45 $\pm$ 0.24                    \\ \hline
        AV2Wav-23-long re-syn        & Target          &    -0.06 $\pm$ 0.22             \\ \bottomrule

\end{tabular}
\end{center}
\end{table}
\begin{table}[h]
\caption{
WER \kledit{(\%)} \klremove{analysis} for each interferer \kledit{type} (speech, noise) and SNR range\klremove{s}. \jcremove{AVSE-}Baseline 
+ AV2Wav denotes \kledit{that} the speech is processed by the baseline first, then re-synthesized using AV2Wav. \jcremove{\jcedit{The SNR range\kledit{s are} \klremove{is} selected based on the pilot evaluation in~\cite{blanco2023avse}.}}}
\label{tab:error}
\resizebox{\columnwidth}{!}{
\begin{tabular}{l|rr|rr|r}
\toprule
   \multicolumn{1}{c|}{\textbf{interferer}}           & \multicolumn{2}{c|}{\textbf{speech}}        &\multicolumn{2}{c|}{\textbf{noise}}     &  Avg.      \\ \hline
     \multicolumn{1}{c|}{\textbf{SNR \kledit{(dB)}}}        & \textbf{[-15,-5]} & \textbf{[-5,5]} & \textbf{[-10, 0]} & \textbf{[0, 10]}  & \\ \midrule
            Mixed speech \kledit{(input)} & 102.4 & 64.4 &24.8  &7.6 & 48.4 \\ \hline
             Baseline~\cite{blanco2023avse}  &   40.3       &    24.1       & 30.6 &  11.6   &  26.4 \\ \hline
             VisualVoice~\cite{gao2021visualvoice}  &   38.9       &    22.8         & N/A &  N/A   &  N/A \\ \hline
             AV2Wav-23-long  &   43.4         &    12.0       & 11.6 &  \textbf{4.0}  &  17.2 \\ \hline
             AV2Wav-23-long  &   105.0         &    66.7       & 26.7 &  6.7  &  49.9 \\ 
             with audio-only input  & & & & \\ \hline
             AV2Wav-23-long-avse  &   43.0         &    11.7       & \textbf{9.8} &  5.3  &  16.8 \\ \hline
             Baseline~\cite{blanco2023avse}  
             &   \textbf{19.8}         &    11.4       & 17.2 & 7.4 & 13.9      \\ 
             + AV2Wav-23-long-avse  &  & & & \\ \hline
             AV2Wav-23-long-avse  &   21.7         &    \textbf{9.7}       & 14.6 &  5.8  &  \textbf{12.8} \\ 
             (fine-tune AV-HuBERT) & & & & & \\ \hline
              Baseline   &   29.3         &    12.4       & 26.1 &  10.2  &  19.4 \\ 
              + AV2Wav-23-long-avse & & & & & \\
             (fine-tune AV-HuBERT) & & & & & \\
             \bottomrule
\end{tabular}
}
\end{table}
\subsubsection{The effect of data filtering}
\vspace{-.05in}
To \klremove{understand the efficacy}\kledit{study the effect} of data filtering using NQE, we compare the following models: 
(1) \textit{AV2Wav-23}: \klremovecr{model }trained on the filtered subset with P-SI-SDR \klremovecr{above }$>23$ (616 hours) (2) \textit{AV2Wav-23-long}: same as (1) but \kleditcr{trained} for 2M steps with a batch size of $64$ 
(3) \textit{AV2Wav-25}: model trained on the filtered subset with P-SI-SDR \klremovecr{above }$>25$ (306 hours) (4) \textit{AV2Wav-random}: \kleditcr{same as (3), but trained on a randomly sampled} 306 hours from \kleditcr{the training set.}\klremovecr{ LRS3 + VoxCeleb2.} \klremovecr{We keep the data size \jcedit{of \textit{AV2Wav-random} }similar to \textit{AV2Wav-25} to \klremove{understand}\kledit{isolate} the effectiveness of filtering.} The objective \kleditcr{and subjective} evaluation results can be found in \kleditcr{Tables~\ref{tab:mixed} and~\ref{tab:cmos}, respectively.}\klremovecr{subjective evaluation can be found in Table~\ref{tab:cmos}. }
\klremove{We can see that }\textit{AV2Wav-random} (\kledit{Table~\ref{tab:mixed} line} \textbf{11}) has a much lower P-SI-SDR than \textit{AV2-Wav-25} (\textbf{10}) on mixed speech\kleditcr{, providing support for NQE-based filtering}. 
\subsubsection{The importance of visual cues}
\jceditar{One natural question to ask is how much improvement we can get from adding visual cues. When comparing We compare AV2Wav-23-long and AV2Wav-23-long with audio-only input (conditioning on $f^a_{1:L}$ in eq.~\ref{eq:f_a_v}) in Table~\ref{tab:error}, and find that AV2Wav is not able to improve over the input mixed speech without access to the visual cues. For low SNR noise interferers, AV2Wav without visual cues performs significantly worse than that with visual cues. For high SNR noise interferers, with visual cues, it still performs better than that without visual cues. It shows that visual cues are useful speech enhancement in the AV2Wav framework. }

\subsubsection{Fine-tuning on AVSE / VCTK}
\jcedit{We fine-tune\klremove{d} the waveform synthesizer on AVSE or VCTK. By default, the AV-HuBERT model is frozen. The objective evaluation can be found in Table~\ref{tab:mixed}.}
We can see that fine-tuning on AVSE (\textit{AV2Wav-23-avse} (\kleditcr{line }\textbf{12})) or VCTK (\textit{AV2Wav-23-vctk} (\textbf{15})) provides some improvement on WER and P-SI-SDR (comparing to \textit{AV2Wav-23} line \textbf{8})\jcremove{, while AVSE provides more improvement than VCTK}. \mm{removed}\jcremove{However, when fine-tuning the model on both datasets, we see a smaller improvement than fine-tuning on one of the datasets. }
\jcremove{The reason could be that the two datasets have very different distributions, so training on them together does not provide additional improvement.  }
For the subjective experiments in Table~\ref{tab:cmos}, after fine-tuning on AVSE (\textit{AV2Wav-23-long-avse} (\textbf{13})), the CMOS improves slightly over the model trained solely on the near-clean subset (\textit{AV2Wav-23-long} (\textbf{9})). 

\jceditar{We also fine-tune the noise-robust AV-HuBERT together with the waveform synthesizer (line \textbf{14} in Table~\ref{tab:mixed} and AV2Wav-23-long-avse (fine-tune AV-HuBERT) in Table~\ref{tab:error}), which may improve performance by exposing AV-HuBERT to AVSE data.  Indeed, the overall WER decreases significantly in this setting, although in the case of low-SNR noise interferers the WER increases.  
From informal listening, fine-tuning AV-HuBERT sometimes results in muffled words, which we hypothesize could be the cause of the WER increase. }

\vspace{-.075in}
\subsubsection{Comparing to the masking-based baselines}
\vspace{-.05in}
In Table~\ref{tab:mixed}, \textit{AV2Wav-23-long-avse} (line \textbf{13}) outperforms the baseline (\textbf{7}) and the original mixed input (\textbf{6}) in terms of WER and P-SI-SDR. \kleditcr{In the subjective evaluation (Table~\ref{tab:cmos}), \textit{AV2Wav-23-long-avse} outperforms the baseline by a large margin.}

\mm{add visualvoice} \kleditcr{In Table~\ref{tab:error} we compare our model to the masking-based baseline\klremovecr{ from the AVSE challenge~\cite{blanco2023avse}} and VisualVoice for source separation~\cite{gao2021visualvoice} in different SNR ranges.}  Our model \jcedit{outperforms} the baseline for \kledit{most} speech and noise interferers. It is \klremove{only}slightly worse than the baseline \kleditcr{for low-SNR speech interferers}. In such cases, the AV-HuBERT model \klremove{could}\kledit{can} not recognize the \klremovecr{corresponding}\kleditcr{target} speech from the mixed speech and the lip motion sequence. \jceditar{However, Fine-tuning AV-HuBERT jointly with the synthesizer helps in addressing this issue. }

We \kledit{also} find that our approach \kledit{combines well with the masking-based baseline\kleditcr{:}} 
By first applying the baseline and then re-synthesizing the waveform using AV2Wav \kledit{given the output from the baseline}, we observe an improvement \klremove{in}\kledit{for} speech interferers\kledit{, especially} at lower SNR\kledit{, over either model alone}. However, fine-tuning AV-HuBERT jointly can achieve lower WER than baseline + AV2Wav.\klremove{In this case, the baseline eliminates some of the noise and our model can \kledit{then} synthesize \kledit{higher-quality}\klremove{the} speech\klremove{ after the baseline}.} 
\klremove{It shows that our model can complement with the masking baseline approach to synthesize better-quality speech.}

\vspace{-.1in}
\subsubsection{Comparing audio quality to target speech}
\vspace{-.05in}
From the target re-synthesis experiments in Table~\ref{tab:mixed}, we can see that \klremove{all of} the re-synthesized speech (line \textbf{2-4}) \kledit{is generally more intelligible (has lower WERs) than}\klremove{has lower WERs than those of} the target speech\kleditcr{ (line \textbf{1})}, \kledit{while maintaining similar estimated audio quality (similar P-SI-SDR)}.\jcremove{ The reason }\klremove{that WER is lower for re-synthesized speech}\jcremove{ could be the extra silence before or after the speech.}
From the subjective evaluation \jcedit{(Table~\ref{tab:cmos})}, the re-synthesized speech (\textit{AV2Wav-23-long re-syn}) is also on par with the original target\kledit{ in terms of CMOS}. Both show that AV2Wav can re-synthesize natural-sounding \jcremove{and high-quality }speech\klremove{ to resemble real speech}. \klremovecr{However, w}\kleditcr{W}hen comparing the enhanced speech\jcedit{ (AV2Wav-23-long-avse)} with the target speech \jcedit{(target)}\jcedit{ in the listening test (Table~\ref{tab:cmos})}, our model is \kleditcr{close but} \klremovecr{still} slightly worse\kleditcr{ (CMOS = -0.45)}. 
\vspace{-.1in}
\subsubsection{Fast inference}
\mm{add line number}
A major disadvantage of diffusion models is their slow inference\klremovecr{, which makes them difficult to use for real-time applications}. As we train the model using continuous noise levels, we can use fewer steps at different noise levels as in~\cite{chen2020wavegrad} (line \textbf{16}). Empirically, we \kledit{find} that $100$ steps can provide good quality speech. We also compare DDIM~\cite{song2020denoising} (line \textbf{17-19}), a\jcremove{n alternative} sampling algorithm for diffusion models\kledit{ that uses fewer steps of non-Markovian inference\klremove{ process}}.  We can see that the WER \jcremove{slightly increases}\jcedit{is similar}\kleditcr{, while P-SI-SDR is worse,} when using the fast inference algorithm\kleditcr{s} (line \textbf{15-17})\kleditcr{ compared to the larger number of inference steps (\textit{AV2Wav-23-long-avse} line \textbf{13})}. \kl{"the fast inference algorithm" = "DDIM"?  which line(s) is line 9 being compared to?}\jc{all}\kl{but most of them have lower WERs}\klremovecr{The P-SI-SDR for fast inference is \klremovecr{also }worse than \kledit{with} 1000 inference steps (line \textbf{13}).} \jcedit{From }\jcremove{With some }informal listening, we \klremove{found}\kledit{find} that \textit{AV2Wav-cont-100} tends to miss some words \klremove{and}\kledit{while}\textit{ AV2Wav-ddim} tends to \klremove{have}\kledit{produce} some white \kleditcr{background} noise\klremovecr{ in the background}.  

\vspace{-.1in}
\section{Conclusion\klremove{ and Future Work}}
\vspace{-.05in}
\klremove{We presented }AV2Wav\klremove{,}\kledit{ is} a simple framework \kledit{for AVSE} based on noise-robust AV-HuBERT\klremove{,} and a diffusion waveform synthesizer\klremove{ to generate the waveform}. \klremove{We show that b}\kledit{B}y training on a \jcremove{filtered }subset of relatively clean speech, along with noise-robust AV-HuBERT, \klremovecr{our AV2Wav model}\kleditcr{AV2Wav} learns to perform speech enhancement without explicitly training it to de-noise. \jceditar{Further fine-tuning the model on clean/noisy pairs further improves its performance. }\klremove{We show that o}\kledit{O}ur model outperforms a masking-based baseline \kledit{in a human listening test}\kleditcr{, and comes close in quality to the target speech.  Potential directions for improvement include \jceditar{futher noise-robust training of AV2Wav on a larger-scale dataset, extension to other languages, }\jcremove{fine-tuning AV-HuBERT (which may help in cases where the model fails, e.g. with low-SNR speech interferers) }and additional work on fast inference}\jcremove{ and is capable of synthesizing }\jcremove{high-quality natural-sounding speech}.
\klremovecr{\mm{thinking about removing the following}

\jcremove{One setting where}\klremove{The model}AV2Wav performs worse \kledit{than the masking-based baseline is with}\klremove{on} low SNR speech interferers\kledit{,} when the pre-trained AV-HuBERT \kledit{may fail}\klremove{ed} 
 to identify the corresponding speech, suggesting that further fine-tuning AV-HuBERT using noise-robust training\jcedit{ on the target domain} might further boost the performance. }

\kl{check the references for arXiv papers that have since been published in a peer-reviewed venue, and also formatting consistency, capitalization, etc.  It is much easier to make formatting consistent if you use string macros for conference/journal names.  For capitalization, use braces as needed, e.g. "\{AVSE\}"}

\section{Acknowledgement}
This work is partially supported by AFOSR grant FA9550-18-1-0166.
\vfill\pagebreak

\label{sec:refs}
\bibliographystyle{IEEEbib.bst}
\bibliography{refs.bib}

\begin{thebibliography}{10}

\bibitem{hou2018audio}
Jen-Cheng~Hou et.al.,
\newblock ``Audio-visual speech enhancement using multimodal deep convolutional neural networks,''
\newblock {\em IEEE Transactions on Emerging Topics in Computational Intelligence}, 2018.

\bibitem{chern2023audio}
I-Chun~Chern et~al.,
\newblock ``Audio-visual speech enhancement and separation by utilizing multi-modal self-supervised embeddings,''
\newblock in {\em IEEE ICASSP Workshops (ICASSPW)}, 2023.

\bibitem{afouras2018conversation}
Triantafyllos~Afouras et.al.,
\newblock ``The conversation: Deep audio-visual speech enhancement,''
\newblock {\em Interspeech}, 2018.

\bibitem{gao2021visualvoice}
Ruohan Gao and Kristen Grauman,
\newblock ``{VisualVoice}: Audio-visual speech separation with cross-modal consistency,''
\newblock in {\em CVPR}, 2021.

\bibitem{yang2022audio}
Karren~Yang et~al.,
\newblock ``Audio-visual speech codecs: Rethinking audio-visual speech enhancement by re-synthesis,''
\newblock in {\em CVPR}, 2022.

\bibitem{hsu2023revise}
Wei-Ning~Hsu et~al.,
\newblock ``{ReVISE}: Self-supervised speech resynthesis with visual input for universal and generalized speech regeneration,''
\newblock in {\em CVPR}, 2023.

\bibitem{pascual2017segan}
Santiago Pascual, Antonio Bonafonte, and Joan Serra,
\newblock ``Segan: Speech enhancement generative adversarial network,''
\newblock {\em arXiv preprint arXiv:1703.09452}, 2017.

\bibitem{richter2023speech}
Julius Richter, Simon Welker, Jean-Marie Lemercier, Bunlong Lay, and Timo Gerkmann,
\newblock ``Speech enhancement and dereverberation with diffusion-based generative models,''
\newblock {\em IEEE/ACM Transactions on Audio, Speech, and Language Processing}, 2023.

\bibitem{serra2022universal}
Joan Serr{\`a}, Santiago Pascual, Jordi Pons, R~Oguz Araz, and Davide Scaini,
\newblock ``Universal speech enhancement with score-based diffusion,''
\newblock {\em arXiv preprint arXiv:2206.03065}, 2022.

\bibitem{polyak21_interspeech}
Adam~Polyak et.al.,
\newblock ``{Speech Resynthesis from Discrete Disentangled Self-Supervised Representations},''
\newblock in {\em Interspeech}, 2021.

\bibitem{cooke2006audio}
Martin~Cooke et.al.,
\newblock ``An audio-visual corpus for speech perception and automatic speech recognition,''
\newblock {\em The Journal of the Acoustical Society of America}, 2006.

\bibitem{chung2018voxceleb2}
Joon~Son Chung, Arsha Nagrani, and Andrew Zisserman,
\newblock ``{VoxCeleb2}: Deep speaker recognition,''
\newblock {\em Interspeech}, 2018.

\bibitem{afouras2018lrs3}
Triantafyllos Afouras, Joon~Son Chung, and Andrew Zisserman,
\newblock ``{LRS3-TED}: a large-scale dataset for visual speech recognition,''
\newblock {\em arXiv preprint arXiv:1809.00496}, 2018.

\bibitem{shi2021learning}
Bowen~Shi et~al.,
\newblock ``Learning audio-visual speech representation by masked multimodal cluster prediction,''
\newblock in {\em ICLR}, 2021.

\bibitem{chen2020wavegrad}
Nanxin~Chen et~al,
\newblock ``{WaveGrad}: Estimating gradients for waveform generation,''
\newblock in {\em ICLR}, 2020.

\bibitem{irvin2023self}
Bryce~Irvin et.al.,
\newblock ``Self-supervised learning for speech enhancement through synthesis,''
\newblock in {\em ICASSP}, 2023.

\bibitem{hung2022boosting}
Kuo-Hsuan~Hung et.al.,
\newblock ``Boosting self-supervised embeddings for speech enhancement,''
\newblock {\em Interspeech}, 2022.

\bibitem{huang2022investigating}
Zili~Huang et.al.,
\newblock ``Investigating self-supervised learning for speech enhancement and separation,''
\newblock in {\em ICASSP}, 2022.

\bibitem{richter2023audio}
Julius Richter, Simone Frintrop, and Timo Gerkmann,
\newblock ``Audio-visual speech enhancement with score-based generative models,''
\newblock {\em arXiv preprint arXiv:2306.01432}, 2023.

\bibitem{shi22_interspeech}
Bowen Shi, Wei-Ning Hsu, and Abdelrahman Mohamed,
\newblock ``{Robust Self-Supervised Audio-Visual Speech Recognition},''
\newblock in {\em Interspeech}, 2022.

\bibitem{neverova2015moddrop}
Natalia~Neverova et~al.,
\newblock ``{ModDrop}: adaptive multi-modal gesture recognition,''
\newblock {\em IEEE Transactions on Pattern Analysis and Machine Intelligence}, 2015.

\bibitem{le2019sdr}
Jonathan Le~Roux et.al.,
\newblock ``{SDR}--half-baked or well done?,''
\newblock in {\em ICASSP}, 2019.

\bibitem{kumar2023torchaudio}
Anurag~Kumar et~al.,
\newblock ``{TorchAudio-Squim}: Reference-less speech quality and intelligibility measures in torchaudio,''
\newblock in {\em ICASSP}, 2023.

\bibitem{radford2023robust}
Alec~Radford et~al.,
\newblock ``Robust speech recognition via large-scale weak supervision,''
\newblock in {\em ICML}, 2023.

\bibitem{blanco2023avse}
Andrea Lorena Aldana~Blanco et~al.,
\newblock ``{AVSE} challenge: Audio-visual speech enhancement challenge,''
\newblock in {\em IEEE Spoken Language Technology Workshop (SLT)}, 2023.

\bibitem{ba2016layer}
Jimmy~Ba et~al.,
\newblock ``Layer normalization,''
\newblock {\em arXiv preprint arXiv:1607.06450}, 2016.

\bibitem{kingma2014adam}
Diederik~P Kingma and Jimmy Ba,
\newblock ``Adam: A method for stochastic optimization,''
\newblock {\em ICLR}, 2015.

\bibitem{veaux2017cstr}
Christophe Veaux, Junichi Yamagishi, Kirsten MacDonald, et~al.,
\newblock ``{CSTR VCTK} corpus: English multi-speaker corpus for cstr voice cloning toolkit,''
\newblock {\em University of Edinburgh. The Centre for Speech Technology Research (CSTR)}, 2017.

\bibitem{loizou2011speech}
Philipos~C Loizou,
\newblock ``Speech quality assessment,''
\newblock in {\em Multimedia analysis, processing and communications}, pp. 623--654. 2011.

\bibitem{song2020denoising}
Jiaming Song, Chenlin Meng, and Stefano Ermon,
\newblock ``Denoising diffusion implicit models,''
\newblock in {\em ICLR}, 2020.

\end{thebibliography}

\end{document}